\documentclass[oneside]{elsart}
\usepackage{amsmath,amsfonts,amssymb,amsbsy,amstext}
\usepackage{graphicx,float}
\journal{Computers \& Fluids}

\newcommand{\dd}{\ensuremath{\mathrm{d}}}

\begin{document}

\begin{frontmatter}

\title{Numerical dissipation and \\
  the bottleneck effect in simulations of\\
  compressible isotropic turbulence}

\author[mpa,wuerzburg]{Wolfram Schmidt\corauthref{cor}}
\ead{schmidt@astro.uni-wuerzburg.de}
\corauth[cor]{Corresponding author.}
\author[mpa]{Wolfgang Hillebrandt}
\author[wuerzburg]{Jens C. Niemeyer}

\address[mpa]{Max-Planck-Institut f\"{u}r Astrophysik,
  Karl-Schwarzschild-Str.\ 1,\\ D-85741 Garching, Germany}
\address[wuerzburg]{Lehrstuhl f\"{u}r Astronomie, Universit\"{a}t
  W\"{u}rzburg, Am Hubland,\\ D-97074 W\"{u}rzburg, Germany}

\begin{abstract}
  The piece-wise parabolic method (PPM) is applied to simulations of
  forced isotropic turbulence with Mach numbers $\sim 0.1\ldots 1$. The
  equation of state is dominated by the Fermi pressure of an
  electron-degenerate fluid. The dissipation in these simulations is
  of purely numerical origin.  For the dimensionless mean rate of
  dissipation, we find values in agreement with known results from
  mostly incompressible turbulence simulations. The calculation of a
  Smagorinsky length corresponding to the rate of numerical
  dissipation supports the notion of the PPM supplying an implicit
  subgrid scale model. In the turbulence energy spectra of various
  flow realisations, we find the so-called bottleneck phenomenon,
  i.~e., a flattening of the spectrum function near the wavenumber of
  maximal dissipation. The shape of the bottleneck peak in the compensated
  spectrum functions is comparable to what is found in turbulence
  simulations with hyperviscosity.  Although the bottleneck effect
  reduces the range of nearly inertial length scales considerably, we
  are able to estimate the value of the Kolmogorov constant. For
  steady turbulence with a balance between energy injection and
  dissipation, it appears that $C\approx 1.7$. However, a smaller
  value is found in the case of transonic turbulence with a large
  fraction of compressive components in the driving force.  Moreover,
  we discuss length scales related to the dissipation, in particular,
  an effective numerical length scale $\Delta_{\mathrm{eff}}$, which
  can be regarded as the characteristic smoothing length of the
  implicit filter associated with the PPM.
\end{abstract}

\end{frontmatter}

\section{Introduction}

The paradigm of statistically stationary isotropic turbulence put
forward by Kolmogorov is based on the balance between energy injection
and dissipation in equilibrium, regardless of the mechanism
of dissipation. This implies the well-known scaling relation
\cite{Kolmog41}
\begin{equation}
  \label{eq:kolmogrv}
  E(k)=C\langle\epsilon\rangle^{2/3}k^{-5/3}
\end{equation}
for the turbulence energy spectrum in the inertial subrange of wave
numbers $k$.  The universality of Kolmogorov scaling was confirmed in
many laboratory experiments and three-dimensional numerical
simulations
\cite{SheJack93,CaoChen96,YeuZhou97,PortPou94,PortWood98,GotFu01,KanIshi03,DobHau03}.
The mean rate of dissipation $\langle\epsilon\rangle$ is determined
by integral length and velocity scales, $L$ and $V$,
respectively, which are related to the properties of energy injection
into the flow.  Normalising $\langle\epsilon\rangle$ in terms of these
characteristic scales, a dimensionless quantity is obtained:
\begin{equation}
  \label{eq:eps_tilde}
  \langle\tilde{\epsilon}\rangle =
  \frac{L}{V^{3}}\langle\epsilon\rangle
\end{equation}
It is customary to specify the dimensionless rate of dissipation in
terms of the similarity parameter $C_{\epsilon}\sim
\langle\tilde{\epsilon}\rangle$. The exact definition of
$C_{\epsilon}$ is given in section~\ref{sc:rate_diss}.  Numerous
attempts have been made to infer the value of $C_{\epsilon}$ either
from laboratory measurements or numerical simulations of turbulent
flows.  It appears that $\langle\tilde{\epsilon}\rangle$ is generally
of the order unity.  Particularly, a universal value of about $0.5$
for statistically stationary isotropic turbulence at sufficiently high
Reynolds number has emerged over the last decade
\cite{KanIshi03,Sreen98,PearKrog02}.  However, virtually all of the
numerical estimates of $C_{\epsilon}$ have been obtained from
simulations of incompressible turbulence so far.  Only very recently,
first results from simulations of compressible flows with higher-order
finite difference methods have become available and, actually, are in
excellent agreement with previously reported values
\cite{PearYou04}. We will extend this approach and present
calculations of the instantaneous rate of dissipation from the energy
budget in numerical simulations of forced compressible turbulence up
to Mach numbers of the order unity. Moreover, the whole range of
developing, steady and decaying turbulence is investigated. In order
to make parameter studies feasible, we chose a rather moderate
resolution of $432^{3}$.

The simulations we have performed also differ in other aspects from
those cited above. Firstly, this work was motivated by the
astrophysical problem of turbulent deflagration in degenerate
stars. Thus, both the equation of state and the absolute values of
physical quantities differ greatly from anything that is encountered
in engineering or in the atmospheric sciences
\cite{HilleNie00,Schmidt04}. In terms of dimensionless quantities,
however, no significant deviations from turbulence in ideal gas are
found. Apart from that, we use a high-order Godunov scheme, namely,
the piece-wise parabolic method (PPM), which applies to fully
compressible flows and, in fact, introduces \emph{numerical
dissipation} \cite{CoWood84}. Of course, the question arises, whether
the action of numerical dissipation properly mimics the effect of
Navier-Stokes viscous dissipation in direct numerical simulations
(DNS) or the action of subgrid scale turbulence stresses in large-eddy
simulations (LES) \cite{SyPort00,RidDri02}. One cannot reasonably
expect that the local rate of dissipation in the former case or the
energy transfer toward unresolved scale in the latter case is exactly
matched by the dissipative effects of the PPM.  However, at
sufficiently high resolution, a nearly inertial range of scales will
emerge, in which the flow is dominated by the dynamics of the
turbulence cascade and, hence, more or less independent of the actual
dissipation mechanism. In addition, if statistical properties such as
the value of $C_{\epsilon}$ or the Kolmogorov constant $C$ were
reproduced correctly in simulations of turbulence with the PPM, then
the evolution of mean quantities and basic structural features of the
computed flow should be in accordance with more accurate methods with
an explicit treatment of dissipation.  Indeed, there are several
numerical results in support of this point of view
\cite{YeuZhou97,PortWood98,HauBrand04}, and we will attempt to
strengthen the case for the PPM in this article.

Particularly, we computed turbulence energy spectrum functions from
selected flow realisations. The spectra for fully developed turbulence
reveal a pile-up of kinetic energy near the wave number of maximum
dissipation.  This is known under the name \emph{bottleneck effect}
and was found in numerous numerical simulations
\cite{SyPort00,KanIshi03,DobHau03}.  The bottleneck is believed to be
a genuine feature of isotropic turbulence and can be attributed to the
partial suppression of non-linear turbulent interactions under the
influence of viscous dissipation \cite{Falk94}. Based on the
turbulence energy spectra, we propose the definition of a
characteristic length scale $\Delta_{\mathrm{eff}}$ of a presumed
implicit filter which is equivalent to smoothing effect of the PPM. If
the equations of motion are solved with spectral methods, then
$\Delta_{\mathrm{eff}}=\Delta$, where $\Delta$ is the length scale of
the numerical grid in physical space. For finite volume methods,
however, the smoothing of velocity fluctuations due to numerical
dissipation on the smallest resolved length scales $l\sim \Delta$
implies $\Delta_{\mathrm{eff}}=\beta\Delta$, where $\beta$ is expected
to assume a valuer larger than unity. Furthermore, values for the
dimensionless rate of dissipation~\ref{eq:eps_tilde} were calculated
from the growth rate of the internal energy corrected for
compressibility effects. Combining values of the numerical dissipation
rate and the effective length scale $\Delta_{\mathrm{eff}}$,
respectively, it was possible to calculate the Smagorinsky constant,
assuming the the numerical dissipation would be statistically
equivalent to the action of a Smagorinsky subgrid scale model.

In section~\ref{sc:isoturb}, we will outline the forcing scheme, the simulation
parameters and the equation of state in our simulations.   The
evaluation of the rate of dissipation is presented in
section~\ref{sc:rate_diss}, the computation of turbulence energy
spectra in section~\ref{sc:spectrum} and, finally, dissipation length
scales are discussed in section~\ref{sc:diss_length}.

\section{Forced isotropic turbulence}
\label{sc:isoturb}

The dynamics of forced turbulence in compressible fluids is determined
by the following set of conservation laws in combination with the
equation of state \cite{LanLif,Warsi}:
\begin{align}
  \label{eq:isoturb_cont}
  \frac{\partial}{\partial t}\rho + 
    \frac{\partial}{\partial x_{i}}\rho v_{i} & = 0,\\
  \label{eq:isoturb_momt}
  \frac{\partial}{\partial t}\rho v_{i} +
    \frac{\partial}{\partial x_{k}}\rho v_{i}v_{k} & =  
    -\frac{\partial}{\partial x_{i}}P + \rho f_{i} + 
    \frac{\partial}{\partial x_{k}}\sigma_{ik}, \\
  \label{eq:isoturb_energy}
  \frac{\partial}{\partial t}E +
    \frac{\partial}{\partial x_{k}}E v_{k} & =
    \rho f_{k}v_{k}.
\end{align}
The field $\vec{f}(\vec{x},t)$ is called the \emph{driving force} and
may be any mechanical force acting upon the fluid and thereby
supplying energy to the flow. The term $\sigma_{ik,k}$ in the momentum
equation is the viscous force per unit volume. In the case of Navier-Stokes
turbulence, the viscous stress tensor $\sigma_{ik}$ is proportional to
the local strain of the velocity field:
\begin{equation}
  \sigma_{ik} = 2\rho\nu S_{ik}^{*} \equiv
  2\rho\nu\left(S_{\!ik} - \frac{1}{3}v_{j,j}\delta_{ik}\right),
\end{equation}
where $S_{ik}=\frac{1}{2}(v_{i,k}+v_{k,i})$.
The coefficient $\nu$ is the \emph{microscopic viscosity} of the
fluid.

According to the Landau criterion \cite{LanLif}, the range of
dynamically relevant length scales in a turbulent flow is of the order
$L/\eta_{\mathrm{K}}\sim \mathrm{Re}^{3/4}$, where $L$ is the
\emph{integral length scale} associated with the spatial variation of
the driving force, and $\eta_{K}$, which is called the
\emph{Kolmogorov scale}, is a characteristic length scale associated
with viscous dissipation. If the Reynolds number $\mathrm{Re}=LV/\nu$
for a flow of characteristic velocity $V$ is large enough, the range
of dynamically relevant length scales will become computationally
intractable. Then the common approach is to run a LES which resolves
only the largest scales and invokes some model for turbulence on
smaller scales.  On the other hand, it has been suggested to let
merely the dissipative effect of a finite-volume scheme smooth out the
flow on length scales just above the numerical cutoff and dispose of
accumulating kinetic energy \cite{RidDri02}.  In particular, this
assumption was put forward and exhaustively tested for the
\emph{piece-wise parabolic method} (PPM)
\cite{CoWood84,PortPou94,PortWood98,SyPort00}.

We chose to follow this approach and adopted the PPM for
simulations of forced isotropic turbulence. In order to produce 
 turbulence, a random driving force is applied
\cite{EswaPope88}. The evolution of the force field is determined by a
three-component \emph{Ornstein-Uhlenbeck} process 
in spectral space \cite{Pope}, which corresponds to the
following stochastic differential equation (SDE) for the
Fourier transform $\hat{\vec{f}}(\vec{k},t)$:
\begin{equation}
  \label{eq:stirr_mode_evol}
  \dd\hat{\vec{f}}(\vec{k},t) = 
  -\hat{\vec{f}}(\vec{k},t)\frac{\dd t}{T} +
  F_{0}\sum_{jlm}\left(\frac{2\sigma^{2}(\vec{k})}{T}\right)^{1/2}
    \delta(\vec{k}-\vec{k}_{jlm})
    \mathfrak{P}_{\zeta}(\vec{k})\cdot\dd\vec{\mathcal{W}}_{t}.
\end{equation}
The discretisation in physical space induces a discrete spectrum of
modes associated with the wave vectors $\vec{k}_{jlm}$. Gaussian random
increments are introduced by the three-component \emph{Wiener process}
$\vec{\mathcal{W}}_{t}$ and are projected by means of the symmetric
operator
\begin{equation}
  (\mathfrak{P}_{ij})_{\zeta}(\vec{k}) = 
  \zeta \mathfrak{P}_{ij}^{\perp}(\vec{k}) + 
  (1-\zeta)\mathfrak{P}_{ij}^{\parallel}(\vec{k}) =
  \zeta\delta_{ij} + (1-2\zeta)\frac{k_{i}k_{j}}{k^{2}}.
\end{equation}
For the spectral weight $\zeta=1$, the physical force field
becomes purely \emph{solenoidal}, i.~e., divergence-free. Choosing $\zeta <
1$, \emph{dilatational} components are generated which compress or
rarify the fluid.  The spectral profile of the driving force is
determined by the variance $\sigma^{2}(\vec{k})$. We use a symmetric
quadratic function centred at the wave number $k_{0}$. For
$k>2k_{0}$, all modes of the force vanish identically. The integral
length scale $L$ is defined by $L=2\pi/k_{0}$. Once $\sigma(\vec{k})$
is normalised, the asymptotic RMS amplitude of the stochastic driving
force depends on $F_{0}$ and the weight $\zeta$ only:
\begin{equation}
  f_{\mathrm{rms}} \simeq (1 - 2\zeta + 3\zeta^{2})F_{0}.
\end{equation}
The linear drift term in the SDE~(\ref{eq:stirr_mode_evol}) causes
any information about the initial conditions to decay exponentially over
the characteristic time $T$ \cite{Pope}.  Consequently, the computed flow will
evolve towards a statistically stationary state which becomes
asymptotically independent of the initial state. In this regime, the
properties of the flow are determined by the three parameters $L$,
$F_{0}$ and $\zeta$. The force magnitude $F_{0}$ is conveniently
expressed as $F_{0}=V^{2}/L$, where the characteristic velocity $V$ is
about the root mean square velocity in the fully developed flow,
and for the integral time scale we naturally set $T=(L/F_{0})^{1/2}=L/V$.

Given the astrophysical background of our work, we employed the
equation of state (EOS) of a degenerate electron gas.  Degeneracy is a
quantum mechanical phenomenon in the limit of vanishing temperature
which, for example, is encountered in so-called white dwarfs. This
kind of stellar remnant emanates from the burnout of stars comparable
in mass to our Sun \cite{HilleNie00}. In white dwarf matter, the
electrons are not bound to nuclei. While the latter are non-degenerate
and obey the ideal gas law, each electron assumes more or less its
lowest possible energy state. Thus, the electrons follow the
Fermi-Dirac statistics. One can show that the resulting degeneracy
pressure of the electron gas dominates over the residual thermal
pressure of all particles in the interior of a white dwarf
\cite{ShaTeuk}. In this case, the equation of state is approximately
given by $P\propto\rho^{5/3}$ independent of the temperature, provided that the
mass density is less than the critical density $\rho_{\mathrm{c}}\sim
m_{\mathrm{p}}/\lambda_{\mathrm{e}}^{3}\approx 2.9\,\cdot
10^{10}\,\mathrm{kg\,m^{-3}}$, where $m_{\mathrm{p}}$ is the mass of
the proton and $\lambda_{\mathrm{e}}=\hbar/m_{\mathrm{e}}c$ is the
Compton wave length of the electron.  At higher densities, the
electrons become relativistic and the isentropic exponent approaches
$4/3$ \cite{ShaTeuk}.

In fact, a realistic EOS which accounts for contributions from
degenerate electrons, non-degenerate nuclei, photons and pair
production over a wide range of densities and temperatures was used in
the simulations we performed.  This EOS cannot be expressed in
analytical form and must be evaluated numerically \cite{Rein01}. The
following initial conditions were chosen in all but one case:
\begin{align}
  \vec{v}(\vec{x},0) &= 0, \\ 
  \rho(\vec{x},0) &= \rho_{0} =  0.02\frac{m_{\mathrm{p}}}{\lambda_{\mathrm{e}}^{3}}
    \approx 5.805\cdot 10^{8}\,\mathrm{kg\,m^{-3}}, \\
  T(\vec{x},0) &= T_{0} = 0.001\frac{m_{\mathrm{e}}c^{2}}{k_\mathrm{B}}
    \approx 5.929\cdot 10^{6}\,\mathrm{K}.
\end{align}
For the simulation with the lowest Mach number, the initial
mass density is larger by a factor $2.5\cdot 10^{3}$ than the above
value.  In addition to $\rho_{0}$, the spectral weight $\zeta$ of the
driving force, the characteristic velocity $V$ and the integral times
scale $T$ as well as supplementary simulation parameters are listed in
table~\ref{tb:dns}. The spectrum of the driving force is determined by
the wave number $k_{0}=2\pi/L=2\pi\alpha/X$, where $\alpha=3$
is the ratio of the domain size $X$ to the integral length $L$.  We
used a grid of $432^{3}$ cells for each simulation, and an integral
length $L=1.44\cdot 10^{3}\,\mathrm{m}$.  This corresponds to a
numerical resolution of $\Delta=10.0\,\mathrm{m}$.

\begin{table}[bht]
  \begin{center}
    \begin{tabular}{|l|l|c|l|c|c|r|c|}
      \hline
      $\zeta$ & $\rho_{0}\,[\mathrm{kg\,m^{-3}}]$ & $V\,[\mathrm{m\,s^{-1}}] $ & $V/c_{0}$ 
      & T [s] & $t_{\mathrm{d}}/T$ & $t_{\mathrm{f}}$/T & $N_{\Delta t}$ \\[3pt]
      \hline
      1.0  & $1.451\cdot 10^{12}$ & $7.290\cdot 10^{5}$ & 0.084 & $1.98\cdot 10^{-3}$ &     & 2.5  & 5815 \\
      1.0  & $5.805\cdot 10^{8}$  & $7.290\cdot 10^{5}$ & 0.42  & $1.98\cdot 10^{-3}$ & 3.0 & 8.0  & 6343 \\
      0.75 & $5.805\cdot 10^{8}$  & $1.153\cdot 10^{6}$ & 0.66  & $1.25\cdot 10^{-3}$ & 5.0 & 10.0 & 6351 \\
      0.2  & $5.805\cdot 10^{8}$  & $2.430\cdot 10^{6}$ & 1.39  & $5.93\cdot 10^{-4}$ & 5.0 & 10.0 & 3356 \\
      \hline
    \end{tabular}
  \end{center}
  \medskip
  \caption{Chosen values for basic simulation parameters, the time of the onset of decay
    $t_{\mathrm{d}}/T$ and the end of each simulation
    $t_{\mathrm{f}}/T$, as well as the total number of time steps
    $N_{\Delta t}$. }
  \label{tb:dns}
\end{table}

From the hydrodynamical point of view, the EOS for degenerate matter
is remarkable, because the pressure is virtually independent of the
temperature. In consequence, thermodynamics and hydrodynamics are
decoupled. In the case of an ideal gas, on the other hand, both the
temperature and the pressure are gradually rising in a turbulent flow
due to the heat generated by kinetic energy dissipation. Strictly
speaking, turbulence can never be stationary in an ideal gas, because
of the changing relation between fluctuations of the density, pressure
and velocity, respectively, with varying temperature.  Of course,
there are limitations in the case of degenerate matter as well,
because the rising internal energy will eventually break the
degeneracy of the electron gas. This happens once the temperature
becomes comparable to the Fermi temperature, which is $\gtrsim
10^{9}\,\mathrm{K}$. In the case of the simulations discussed
subsequently, the total energy injected into the system is a
significant fraction of the initial internal energy, but degeneracy is
maintained over several integral time scales.

\begin{figure}
  \begin{center}
    \includegraphics[width=\linewidth]{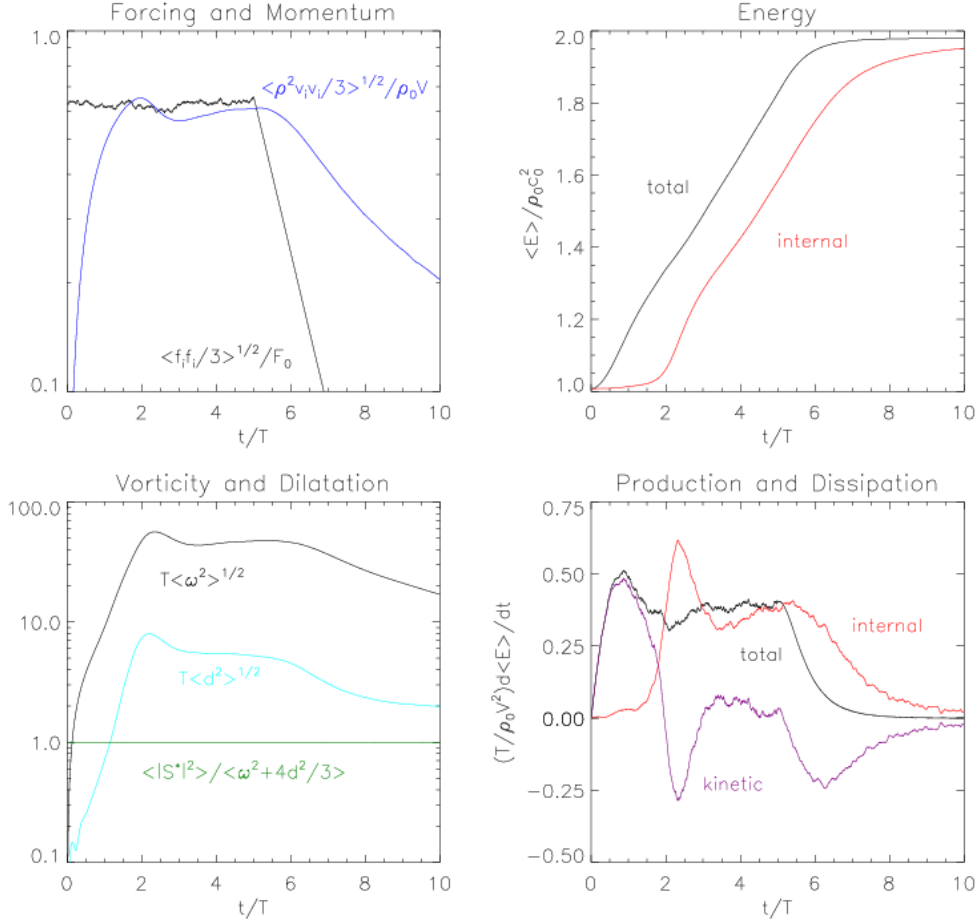}
    \caption{Evolution of dimensionless mean quantities for a
      simulation of forced isotropic turbulence in three dimensions.  The
      characteristic Mach number is $V/c_{0}=0.66$ and the spectral weight
      of the driving force $\zeta=0.75$.  The panels show the RMS momentum
      and specific force, the mean total and internal energy, the RMS
      vorticity and divergence as well as the averaged rates of energy
      production and dissipation as functions of the normalised time
      $\tilde{t}=t/T$.}
    \label{fg:stat3d075}
  \end{center}
\end{figure}

\begin{figure}
  \begin{center}
    \includegraphics[width=\linewidth]{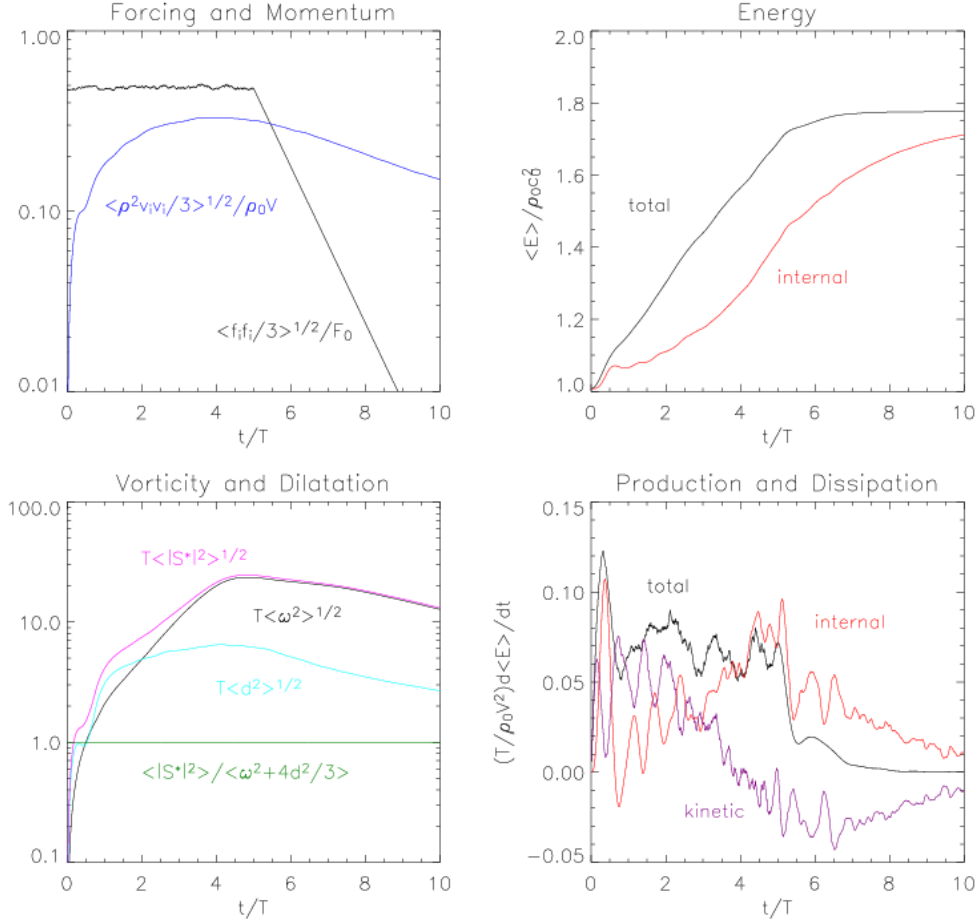}
    \caption{ Statistics for a simulation with $V/c_{0}=0.66$ and $\zeta=0.75$. }
    \label{fg:stat3d020}
  \end{center}
\end{figure}

The global statistics of the simulation with $V\approx 0.66c_{0}$,
where $c_{0}$ is the initial speed of sound, and $\zeta=3/4$ is shown
in figure~\ref{fg:stat3d075}. In this case, the major component of the
driving force is solenoidal, but there is a significant dilatational
contribution as well. The plot of the RMS velocity shown in the top
left panel illustrates the different regimes of the flow evolution. In
the beginning, fluid is gradually set into motion. In the course of
production, the intensity of turbulence grows exponentially, as one
can see from the evolution of the RMS vorticity
$\omega=|\nabla\times\vec{v}|$ in the bottom panel on the left.  At
time $\tilde{t}=t/T\approx 2$, the increase of the kinetic energy
stagnates. From $\tilde{t}\approx 3$ onwards, the mean kinetic energy
remains roughly constant, and there is an approximate balance between
energy injection and dissipation (panels on the right of
figure~\ref{fg:stat3d075}).  Moreover, the graph of the mean vorticity
flattens for $\tilde{t}\gtrsim 2$.  Altogether, we take the evolution
of the statistical quantities as phenomenological indication of
\emph{fully developed turbulence} at time $\tilde{t}\gtrsim 2$, which
asymptotically approaches a \emph{statistically homogeneous} and
\emph{stationary} state.  The conspicuous filaments of intense
vorticity, which are the hallmark of developed turbulence, can be seen
in a three-dimensional visualisation of the flow at time
$\tilde{t}=4.0$ in the figure~\ref{fg:vort3d324}. The intermittency of
turbulence becomes manifest in the spacious voids between the vortex
filaments. At time $\tilde{t}=\tilde{t}_{\mathrm{d}}=5.0$, the
dropping of the random diffusion term in the
SDE~(\ref{eq:stirr_mode_evol}) initiates the exponential decay of the
force field, and the energy contents of the flow is subsequently
dissipated.

Quite a different behaviour emerges in the case $V/c_{0}=1.39$ and
$\zeta=1/5$.  As one can see from the plots in
figure~\ref{fg:stat3d020}, the rise of the RMS velocity is less
steep. Actually, the evolution of the RMS structural invariants
$\langle|S^{\ast}|^{2}\rangle^{1/2}$, $\langle\omega^{2}\rangle^{1/2}$
and $\langle d^{2}\rangle^{1/2}$, which is shown in the left bottom
panels, suggest that there is an initial phase which is dominated by
shocks rather than eddies. This is reflected in pronounced
oscillations in the rate of change of kinetic and internal energy,
respectively, due to compression and rarefaction effects. At
$\tilde{t}\approx 2$, vorticity begins to dominate over the
divergence. Apparently, there is a transition from the shock-dominated
phase toward a regime, in which the flow becomes increasingly
solenoidal, despite of the mostly dilatational force acting upon the
fluid. However, since the decay regime was initiated at the time
$\tilde{t}=5.0$ too, a statistically stationary flow was not
established in this simulation.

Furthermore, we computed a couple of mean structural invariants which
correspond to third-order statistical moments of the rate of strain
tensor $S_{ij}$. Normalisation with respect to the mean rate of strain
$|S|=(2S_{ij}S_{ij})^{1/2}$ gives rise to the definition of the
following parameters:
\begin{align}
  \label{eq:smag}
  a &= \frac{\langle\rho|S^{\ast}|^{2}|S|\rangle}
            {\rho_{0}\langle|S|^{2}\rangle^{3/2}}, \\
  \label{eq:skew}
  b &= \frac{\langle\rho
                S_{\!ij}S_{jk}S_{kl}\rangle}{\rho_{0}\langle|S|^{2}\rangle^{3/2}}
         = \frac{1}{2\sqrt{2}}\mathrm{skew}(S_{\!ik}).
\end{align}
The first parameter, $a$, is closely related to the mean rate of
dissipation in the Smagorinsky model. We will comment on this point
further in the following section. The parameter $b$, on the other
hand, is proportional to the skewness of the rate of strain tensor.
Numerical values of $a$ and $b$ evaluated from flow
realisations in three simulations at certain instants of time are
summarised in table~\ref{tb:skew}.  Thereby, a remarkable similarity
is revealed.  In particular, the rate-of-strain skewness of about
$-0.3$ agrees with known results for isotropic turbulence \cite{Kosov97}.
We take this as a further indication that the flow realisations in our
simulations are in good approximation both statistically stationary
and isotropic after a few integral time scales have elapsed.

\begin{table}[bht]
  \begin{center}
  \begin{tabular}{|l|c|c|c|c|c|c|}
    \hline
    $V/c_{0}$ & $t$ & $a$ & $b$ & $\mathrm{skew}(S_{\!ik})$ &
    $\ell_{\mathrm{S}}/\Delta$ & $\ell_{\mathrm{S}}/\Delta_{\mathrm{eff}}$ \\
    \hline
    0.084 & 2.5 & 1.342 & -0.1045 & -0.296 & 0.298 & 0.17 \\
    \hline
    0.42  & 3.0 & 1.339 & -0.1041 & -0.295 & 0.260 & 0.16 \\
    \hline
    0.66  & 4.0 & 1.333 & -0.1037 & -0.293 & 0.258 & 0.16 \\
    \hline
  \end{tabular}
  \caption{ Normalised third-order moments of the rate of strain
  	defined in equations~(\ref{eq:smag}) and~(\ref{eq:skew}) and the
  	equivalent Smagorinsky length in units of $\Delta$ and $\Delta_{\mathrm{eff}}$,
  	respectively. }
  \label{tb:skew}
  \end{center}
\end{table}

\section{The rate of dissipation}

\label{sc:rate_diss}

The details of the numerical dissipation produced by the PPM are
basically unknown, but it is possible to infer the mean rate of
dissipation from the globally averaged energy conservation laws
\cite{Warsi}. In fact, there are two distinct mechanisms of
dissipation in a compressible fluid. On the one hand,
\emph{pressure-dilatation} accounts for the conversion of internal
energy into mechanical energy and vice versa, as fluid, respectively,
expands or contracts.  Although pressure-dilatation might locally
produce mechanical work, it is effectively a dissipative, irreversible
process, and the net rate of heat production is given by $-\langle P
d\rangle$, where $d=S_{ii}$ is the divergence of the velocity
field. If pronounced shocks are present, however, there might be
transient phases in which $\langle P d\rangle$ becomes positive. This
can be seen, for example, in the right bottom panel of
figure~\ref{fg:stat3d020}, where the time derivative of the internal
energy exhibits local minima corresponding to the production of
mechanical energy by pressure-dilatation. On the other hand, the
change of internal energy caused by \emph{viscous dissipation} is
strictly negative.

For periodic boundary conditions, the total flux through the boundary
surfaces cancels out, and averaging of the internal energy equation
thus yields
\begin{equation}
  \rho_{0}\epsilon_{\mathrm{num}} = 
  \frac{\dd}{\dd t}\langle E_{\mathrm{int}}\rangle + \langle P d\rangle.
\end{equation}
Since we do not employ an explicit viscosity term in the equation of
motion, $\epsilon_{\mathrm{num}}$ is indeed the mean rate of dissipation
due to numerical effects.  The corresponding dimensionless dissipation
rate $\tilde{\epsilon}_{\mathrm{num}}$ is defined by
equation~(\ref{eq:eps_tilde}).  Several representative values of
$\tilde{\epsilon}_{\mathrm{num}}$ and the ratio $-\langle P
d\rangle/\epsilon_{\mathrm{num}}$ 
at certain instants of time are listed in the
tables~\ref{tb:num_nrsl}, \ref{tb:num_nrh75} and~\ref{tb:num_nrh20}.

In the literature, it is common to normalise the rate of dissipation in
terms of the RMS velocity fluctuation $v^{\prime}$ and an integral
length scale $\hat{L}$, which is defined by the transversal turbulence
energy spectrum $E(k)^{\perp}$:
\begin{equation}
  \label{eq:l_hat}
  \hat{L} =
  \frac{\pi}{2v^{\prime\,2}}\int_{0}^{\pi/\Delta}\frac{E(k)^{\perp}\,\dd k}{k}
  \simeq\frac{L}{4v^{\prime\,2}}\sum_{n}\frac{\Phi_{n}^{\perp}}{\tilde{k}_{n}}.
\end{equation}
The average kinetic energy $\Phi_{n}^{\perp}$ contained in Fourier
modes of wave number $k_{n}$ will be defined in the next section.  For
isotropic turbulence, $v^{\prime\,2}=\frac{2}{3}\langle
e_{\mathrm{kin}}\rangle$, where $e_{\mathrm{kin}}=\frac{1}{2}v^{2}$ is
the specific kinetic energy. The parameter $C_{\epsilon}$ specifying
the dimensionless rate of dissipation is then given by
\begin{equation}
  C_{\epsilon} =
  \frac{\hat{L}}{v^{\prime\,3}}\langle\epsilon_{\mathrm{num}}\rangle.
\end{equation}
Using the instantaneous values of the rate of dissipation and the RMS
velocity fluctuations, we estimated $C_{\epsilon}$ for different flow
realisations. The accuracy is limited by the calculation of $\hat{L}$,
which is mostly determined by the smallest wave numbers, and only few
discrete cells are available for $\tilde{k}\sim 1$ in Fourier
space. The results are listed in the tables~\ref{tb:num_nrsl},
\ref{tb:num_nrh75} and~\ref{tb:num_nrh20}. Generally, $C_{\epsilon}$
is small compared told unity in the production phase.  For steady
turbulence, values in the range $0.4\ldots0.5$ are found, which are
close to the time-averaged asymptote $\bar{C}_{\epsilon}\approx 0.5$
in simulations of incompressible turbulence
\cite{KanIshi03,Sreen98,PearKrog02}.

In the introduction, we mentioned that simulations
of Eulerian fluids with the PPM, to a certain degree, are equivalent
to a LES with explicit modelling of the energy transfer toward unresolved
scales in place of numerical dissipation. The consistency of this
assumption in terms of statistical quantities can be corroborated for a 
simple algebraic subgrid scale model such as the Smagorinsky model.
The closure for the rate of dissipation in the Smagorinsky model is given by
\begin{equation}
   \langle\epsilon_{\mathrm{sgs}}\rangle = 
   a\ell_{\mathrm{S}}^{2}\langle|S|^{2}\rangle^{3/2}, \\
\end{equation}
where the coefficient $a$ is defined by equation~\ref{eq:smag}, $|S|$
is the rate of strain and $\ell_{\mathrm{S}}$ is a characteristic
length of the model. Invoking the condition that
$\langle\epsilon_{\mathrm{num}}\rangle$ equals the rate of dissipation
predicted by the Smagorinsky model for the given average rate of
strain in a particular flow realisation, the Smagorinsky length
$\ell_{\mathrm{S}}$ can be calculated. The obtained values in units of
$\Delta$ for developed turbulence of three different characteristic
Mach numbers are listed in table~\ref{tb:skew}. Moreover, we
normalised $\ell_{\mathrm{S}}$ in terms of the effective length scale
$\Delta_{\mathrm{eff}}$.  The procedure for the calculation of
$\Delta_{\mathrm{eff}}$ is introduced in
section~\ref{sc:diss_length}. As we shall argue,
$\Delta_{\mathrm{eff}}$ is the appropriate cutoff scale for the
PPM. Indeed, the results listed in table~\ref{tb:skew} verify in each
case that $\ell_{\mathrm{S}}/\Delta_{\mathrm{eff}}$ is about $0.16$,
which is just the Smagorinsky constant $C_{\mathrm{S}}$ obtained from
analytical considerations \cite{Pope}.

\begin{table}[bht]
  \begin{center}
  \begin{tabular}{|c|c|c|c|c|c|l|c|c|c|}
    \hline
    $\tilde{t}$ & $\langle \tilde{e}_{\mathrm{kin}}\rangle$ &
    $\langle e_{\mathrm{kin}}^{\parallel}\rangle/\langle e_{\mathrm{kin}}\rangle$ &  
    $\tilde{\epsilon}_{\mathrm{num}}$ & 
    $-\langle P d\rangle/\rho_{0}\epsilon_{\mathrm{num}}$ &
    $C_{\epsilon}$ & $\hat{L}/L$ & $l_{\mathrm{p}}/L$ & $\beta$ \\
    \hline
    1.5 & 0.891 & $3.74\cdot 10^{-3}$ & 0.130 & 0.422 & 0.077 & 0.270 &        & 1.82 \\
    2.0 & 0.885 & $6.74\cdot 10^{-3}$ & 0.985 & 0.073 & 0.51  & 0.232 & 0.0659 & 1.57 \\
    3.0 & 0.728 & $8.16\cdot 10^{-3}$ & 0.474 & 0.107 & 0.40  & 0.281 & 0.0674 & 1.62 \\
    4.0 & 0.574 & $9.41\cdot 10^{-3}$ & 0.498 & 0.053 & 0.57  & 0.269 & 0.0615 & 1.61 \\
    6.0 & 0.168 & $2.13\cdot 10^{-2}$ & 0.088 & 0.051 & 0.83  & 0.343 & 0.0644 & 1.69 \\
    \hline
  \end{tabular}
  \medskip
  \caption{Mean energy, rate of dissipation and characteristic length
    scales for a DNS with $V/c_{0}=0.42$ and $\zeta=1.0$.}
  \label{tb:num_nrsl}
  \end{center}
\end{table}

\begin{table}[bht]
  \begin{center}
  \begin{tabular}{|c|c|c|c|c|c|l|c|c|c|}
    \hline
    $\tilde{t}$ & $\langle \tilde{e}_{\mathrm{kin}}\rangle$ &
    $\langle e_{\mathrm{kin}}^{\parallel}\rangle/\langle e_{\mathrm{kin}}\rangle$ &  
    $\tilde{\epsilon}_{\mathrm{num}}$ & 
    $-\langle P d\rangle/\rho_{0}\epsilon_{\mathrm{num}}$ &
    $C_{\epsilon}$ & $\hat{L}/L$ & $l_{\mathrm{p}}/L$ & $\beta$ \\
    \hline
    2.0 & 0.668 & $2.25\cdot 10^{-2}$ & 0.322 & 0.158 & 0.26 & 0.229 & 0.0830 & 1.66 \\
    4.0 & 0.544 & $2.10\cdot 10^{-2}$ & 0.305 & 0.106 & 0.40 & 0.279 & 0.0601 & 1.62 \\
    5.0 & 0.571 & $1.87\cdot 10^{-2}$ & 0.350 & 0.070 & 0.44 & 0.285 & 0.0659 & 1.61 \\
    7.0 & 0.252 & $2.75\cdot 10^{-2}$ & 0.170 & 0.059 & 0.74 & 0.286 & 0.0601 & 1.61 \\
    9.0 & 0.092 & $6.72\cdot 10^{-2}$ & 0.037 &-0.150 & 0.86 & 0.318 & 0.0644 & 1.69 \\    
    \hline
  \end{tabular}
  \medskip
  \caption{Mean energy, rate of dissipation and characteristic length
    scales for a DNS with $V/c_{0}=0.66$ and $\zeta=0.75$.}
  \label{tb:num_nrh75}
  \end{center}
\end{table}

\begin{table}[bht]
  \begin{center}
  \begin{tabular}{|c|c|c|c|c|c|l|c|c|c|}
    \hline
    $\tilde{t}$ & $\langle \tilde{e}_{\mathrm{kin}}\rangle$ &
    $\langle e_{\mathrm{kin}}^{\parallel}\rangle/\langle e_{\mathrm{kin}}\rangle$ &  
    $\tilde{\epsilon}_{\mathrm{num}}$ & 
    $-\langle P d\rangle/\rho_{0}\epsilon_{\mathrm{num}}$ &
    $C_{\epsilon}$ & $\hat{L}/L$ & $l_{\mathrm{p}}/L$ & $\beta$ \\
    \hline
    2.0 &  0.103 & 0.250 & 0.0105 & 0.663 & 0.30 & 0.332 & 0.323 & 5.57 \\
    3.5 &  0.157 & 0.195 & 0.0278 & 0.940 & 0.30 & 0.260 & 0.229 & 2.24 \\
    5.0 &  0.147 & 0.189 & 0.0618 & 0.147 & 0.66 & 0.241 & 0.066 & 1.75 \\
    6.0 &  0.121 & 0.134 & 0.0443 & 0.198 & 0.60 & 0.251 & 0.063 & 1.71 \\
    9.0 &  0.045 & 0.158 & 0.0141 &-0.078 & 0.93 & 0.265 & 0.060 & 1.68 \\
    \hline
  \end{tabular}
  \medskip
  \caption{Mean energy, rate of dissipation and characteristic length
    scales for a DNS with $V/c_{0}=1.39$ and $\zeta=0.2$.}
  \label{tb:num_nrh20}
  \end{center}
\end{table}

\section{Turbulence energy spectra}

\label{sc:spectrum}

In the case of isotropic turbulence, the sum over all
squared Fourier modes of the velocity field can be expressed as an integral
over a function of the wave number only:
\begin{equation}
  \label{eq:energy_spect_intgr}
  \frac{1}{V^{2}}\sum_{jlm}\frac{1}{2}
    \langle\hat{\vec{v}}_{jlm}(t)\cdot\hat{\vec{v}}_{jlm}^{\,\ast}(t)\rangle =
  \int_{0}^{\infty}\dd(\alpha\tilde{k})\tilde{E}(\alpha\tilde{k},t)
\end{equation}
Here the wave number is written in dimensionless form as $\tilde{k}=Lk/2\pi$.
The function $E(k,t)=(\alpha L/2\pi)V^{2}\tilde{E}(\alpha\tilde{k},t)$
is called the \emph{energy spectrum function}. For a discrete spectrum
of modes, $\tilde{E}(\alpha\tilde{k},t)$ is a generalised function
which is defined by:
\begin{equation}
  \tilde{E}(\alpha\tilde{k},t) = \frac{1}{V^{2}}\sum_{jlm}\frac{1}{2}
    \langle\hat{\vec{v}}_{jlm}(t)\cdot\hat{\vec{v}}_{jlm}^{\,\ast}(t)\rangle
    \delta[\alpha\tilde{k}-(j^{2}+l^{2}+m^{2})^{1/2}].
\end{equation}
For the numerical evaluation, we define an approximation to 
$\tilde{E}(\alpha\tilde{k},t)$ by summing over wave number bins
$[k_{n-1/2},k_{n+1/2}]$. This yields the discrete set of numbers
\begin{equation}
  \label{eq:disc_spect}
  \tilde{E}_{n}(t) = \frac{4\pi (\alpha\tilde{k}_{n})^{2}}{V^{2}}\Phi_{n}(t),
\end{equation}
where $k_{n}=\frac{1}{2}(k_{n-1/2}+k_{n+1/2})$, and $\Phi_{n}(t)$ is
the average kinetic energy of all modes in the corresponding wave
number bin:
\begin{equation}
  \label{eq:phi_n}
  \Phi_{n}(t) = \frac{1}{\tilde{\mu}_{n}}
  \sum_{\alpha^{2}\tilde{k}_{n-1/2}^{2}\le\alpha^{2}\tilde{k}^{2}
        \le\alpha^{2}\tilde{k}_{n+1/2}^{2}}
    \sum_{jlm}\frac{1}{2}\hat{\vec{v}}_{jlm}(t)\cdot\hat{\vec{v}}_{jlm}^{\,\ast}(t)
    \delta_{j^{2}+l^{2}+m^{2}}^{\alpha^{2}\tilde{k}^{2}}.
\end{equation}
Here $\tilde{\mu_{n}}$ is the statistical weight of the $n$th
wave number bin, i.~e., the number of cells in Fourier space with wavenumber
$k_{jlm}\in[k_{n-1/2},k_{n+1/2}]$.

\begin{figure}[thb]
  \begin{center}
    \includegraphics[width=\linewidth]{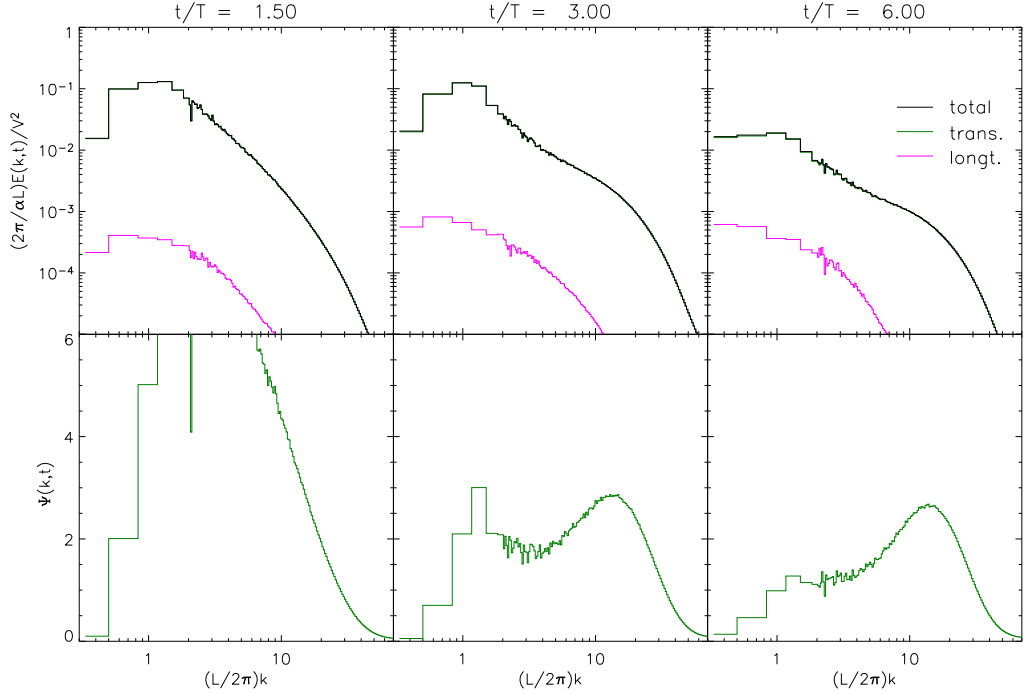}
    \caption{ Turbulence energy spectrum functions for $V/c_{0}=0.42$ and
      $\zeta=1.0$. In the bottom panels, the compensated spectrum
      functions corresponding to the transversal component are plotted. }
    \label{fg:spect100}
  \end{center}
\end{figure}

\begin{figure}[thb]
  \begin{center}
    \includegraphics[width=\linewidth]{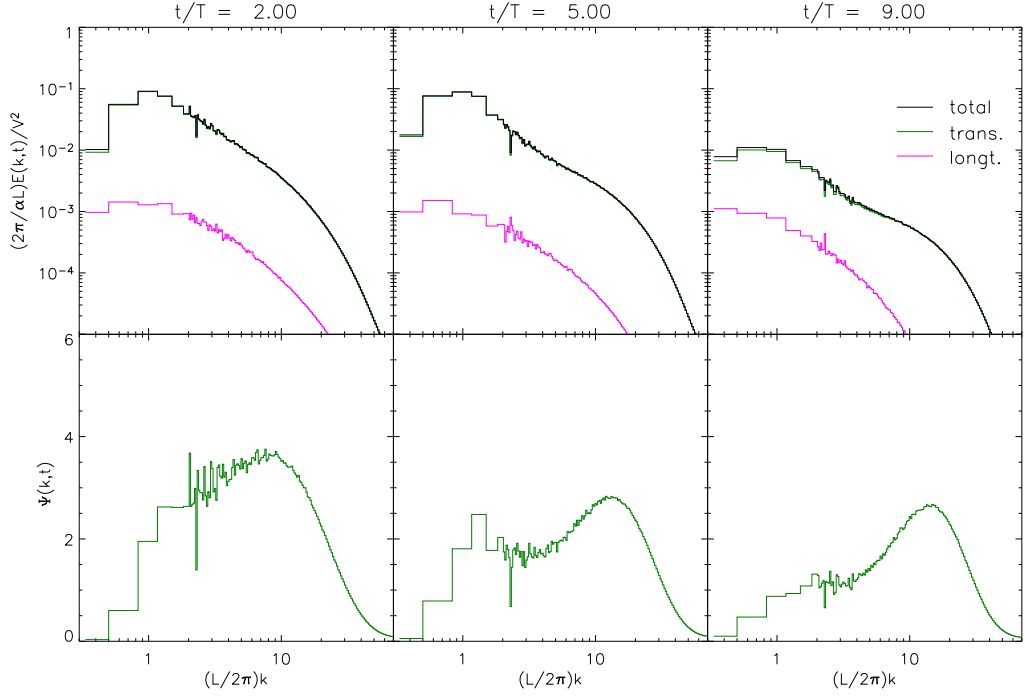}
    \caption{ Turbulence energy spectra for $V/c_{0}=0.66$ and $\zeta=0.75$. }
    \label{fg:spect075}
  \end{center}
\end{figure}

\begin{figure}[thb]
  \begin{center}
    \includegraphics[width=\linewidth]{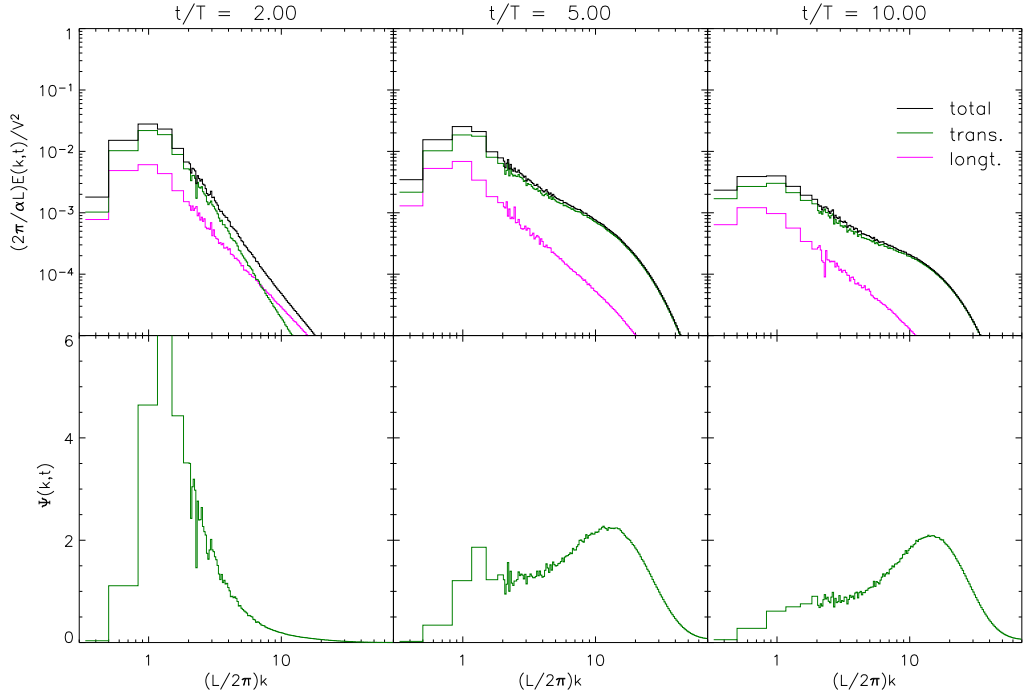}
    \caption{ Turbulence energy spectra for $V/c_{0}=1.39$ and $\zeta=0.2$. }
    \label{fg:spect020}
  \end{center}
\end{figure}

A number of spectra were computed from flow realisations at selected
times. Some of these spectra are shown in Figures~\ref{fg:spect100},
\ref{fg:spect075} and~\ref{fg:spect020}.  For the calculation of the
discrete spectrum function as given by
definition~(\ref{eq:disc_spect}), a coarse equidistant wave number
mesh was used in the energy containing range, $0\le\tilde{k}\le2$, and
a logarithmic mesh with narrow bins for the larger wave numbers.  The
panels on the top of figure~\ref{fg:spect100} show, from left to
right, the discrete energy spectrum function $\tilde{E}_{n}(t)$ at
representative stages for the simulation with $\zeta=1.0$ and
characteristic Mach number $V/c_{0}\approx 0.42$.  The panel on the
very left shows a spectrum in the production phase.  The middle panel
corresponds to developed turbulence, and the panel on the very
right shows a spectrum for decaying turbulence.  One can clearly see
the larger energy contents at high wave numbers in the case of
developed turbulence and the lower overall energy budget in the decay
regime. Also plotted are the longitudinal and transversal spectrum
functions, $\tilde{E}_{n}^{\parallel}(t)$ and
$\tilde{E}_{n}^{\perp}(t)$,
respectively. $\tilde{E}_{n}^{\parallel}(t)$ corresponds to the
dilatational (rotation-free) components of the flow and
$\tilde{E}_{n}^{\perp}(t)$ to the solenoidal (divergence-free)
components. From the plots in figure~\ref{fg:spect100}, it becomes
apparent that the longitudinal energy fraction at higher wave numbers
decays faster in comparison to the transversal fraction. The energy in
the longitudinal components at small wave numbers, on the other hand,
remains almost constant.

From the Kolmogorov spectrum function~(\ref{eq:kolmogrv}), it follows
that $\langle\epsilon\rangle^{-2/3}k^{5/3}E^{\perp}(k)$ is
approximately constant in the inertial subrange of wave numbers. This
suggests the definition of a \emph{compensated spectrum function},
\begin{equation}
  \label{eq:comps_spect}
  \Psi_{n}^{\perp}(t) =
  \left[\frac{\alpha}{2\pi}\langle\tilde{\epsilon}(t)\rangle\right]^{-2/3}
  (\alpha\tilde{k}_{n})^{5/3}\tilde{E}_{n}^{\perp}(t),
\end{equation}
as an indicator of Kolmogorov scaling.  Note that only the transversal
part of the energy spectrum is compensated, because it is the
incompressible fraction of turbulence energy which \emph{a priori}
fulfils Kolmogorov scaling.  For the calculation of
$\Psi_{n}^{\perp}(t)$, the values of
$\langle{\tilde\epsilon}_{\mathrm{num}}\rangle$ listed in the
tables~\ref{tb:num_nrsl}, \ref{tb:num_nrh75} and~\ref{tb:num_nrh20}
were substituted for the dimensionless mean rate of dissipation in
equation~(\ref{eq:comps_spect}).  The resulting plots of the
compensated spectrum functions for $\zeta=1.0$ are shown in the bottom
panels of figure~\ref{fg:spect100}. In particular, the graph of
$\tilde{\Psi}_{n}^{\perp}(t)$ for $t=3.0$ verifies the existence of a
narrow window of wave numbers in the vicinity of $\tilde{k}=3.0$, in
which Kolmogorov scaling with $C\approx 1.7$ applies within the bounds
set by the numerical uncertainty. It is noteworthy that this value is
very close to results obtained from other numerical simulations,
especially, those with higher resolution
\cite{YeuZhou97,GotFu01,KanIshi03}. However, the Kolmogorov constant
inferred from our simulations might be systematically too large,
because of the lack of isotropy for wave numbers close to the
energy-containing subrange \cite{YeuZhou97}.  Moreover, the portion of
the compensated spectrum function $\tilde{\Psi}_{n}^{\perp}(t=3T)$ in
the range of dimensionless wave numbers between $2$ and about $4$
appears to be consistent with a modification of the Kolmogorov
exponent by $-0.1$ , which has recently been inferred
from a DNS with extremely high resolution \cite{KanIshi03}.

Basically the same results hold for the simulation with higher
characteristic Mach number, $V/c_{0}\approx 0.66$, and
$\zeta=0.75$. The sample spectra shown in figure~\ref{fg:spect075} are
very similar in shape compared to those in figure~\ref{fg:spect100},
except for the smaller gap between the total and the longitudinal
energy spectrum functions.  As for the case $\zeta=0.2$ and
$V/c_{0}\approx 1.39$, differences can evidently be seen in
figure~\ref{fg:spect020}.  At early time, the flow is dominated by
large shock waves and most of the energy is contained in longitudinal
modes. This is illustrated by the plot of the energy spectrum
functions at time $\tilde{t}=2.0$. The scaling deviates markedly from
the Kolmogorov law. In particular, the longitudinal component
$E_{n}^{\parallel}(t)$ clearly obeys a power law with exponent $-2$
and the transversal component is falling off even steeper. From
statistics of the velocity field shown in figure~\ref{fg:stat3d020}, a
transition to the regime of solenoidal turbulent flow around
$\tilde{t}\approx 3.0$ can
be discerned \cite{PortPou92}. In this regime, the small-scale
dynamics is eventually dominated by turbulent vortices as in the
case of subsonic turbulence, and the transversal component of the
energy spectrum function is more or less Kolmogorovian. However, we
have a somewhat smaller value of the Kolmogorov constant, $C\approx
1.3$, at time $\tilde{t}=5.0$ (see the middle panels in
figure~\ref{fg:spect020}).

For developed turbulence, there is a pronounced maximum of
the compensated spectrum function at $\tilde{k}\approx 15$
corresponding to a flattening of the energy spectrum in comparison to
the Kolmogorov law. This so-called \emph{bottleneck effect} has
actually been observed in various numerical simulations
\cite{CaoChen96,SyPort00,DobHau03,HauBrand04}. 
The anomalous bottleneck scaling is attributed to dynamical
peculiarities on scales which are significantly influenced by
dissipation.  For example, experimental data indicate a $k^{-1}$ power
law behaviour of the energy spectrum function in the vicinity of the
wave number of maximum dissipation \cite{SheJack93}. A theoretical
explanation of the bottleneck effect on grounds of the non-linear
turbulent transfer was suggested too \cite{Falk94}.  The peak of
$\tilde{\Psi}_{n}(t)$ close to the wave number $\tilde{k}\approx 15$
is more pronounced than in simulations of incompressible turbulence
with spectral methods. For instance, the peak value in the middle panels of the
figures~\ref{fg:spect100} and~\ref{fg:spect075} is about $2.9$,
whereas the peaks of the compensated spectra in
\cite{CaoChen96,KanIshi03} are not much higher than $2.0$. The width of
the bottleneck, on the other hand, is about one decade in wave number
space in all cases.  Remarkably, the properties of the bottleneck we
observe are very similar to what is found in turbulence simulations with
hyperviscosity \cite{HauBrand04}.  Thus, it appears that the numerical
viscosity of the PPM acts like a hyperviscosity, at least as far as
the spectral distribution of turbulence energy is concerned.

\section{Dissipation length scales}
\label{sc:diss_length}

One can regard the action of numerical dissipation as being equivalent
to an \emph{implicit filter} smoothing the flow on a certain length
scale, say, $\Delta_{\mathrm{eff}}$. Fourier modes of wave number
larger than $\pi/\Delta_{\mathrm{eff}}$ are suppressed.
It has already been pointed out in section~\ref{sc:rate_diss} that the ratio
$\beta=\Delta_{\mathrm{eff}}/\Delta$ is of particular interest for
subgrid scale models, which depend on
the reliable specification of the cutoff length.
In combination with a finite-volume method, the cutoff length may
very well be different form the grid resolution $\Delta$.
Although the common point of view holds that it is
not sensible to apply a subgrid scale model in combination with
dissipative schemes such as the PPM, subgrid scale models
are used in some applications without active coupling to the
resolved flow. A particular example is the modelling of turbulent
combustion, where the subgrid scale energy is treated as a
passive scalar and is used as input for a turbulent flame speed
model \cite{NieHille95,HilleNie00,Schmidt04}. 

The determination of the effective length scale $\Delta_{\mathrm{eff}}$ 
for the PPM makes use of the notion of a characteristic filter scale
\cite{Lund97}. For a one-dimensional filter of explicitly known
functional form, the characteristic length can be calculated from
the second moment of the Fourier transform of the filter kernel, the
so-called \emph{transfer function}.  However, since the implicit filter
associated with the PPM is not explicitly known, we have to resort to
the numerically computed energy spectra. Using Kolmogorov's law as
reference spectrum function, we define an \emph{effective
transfer function}
\begin{equation}
  \label{eq:num_flt_transf}
  \hat{G}_{\mathrm{eff}}^{2}(k,t) = 
  \frac{E(k,t)}{E_{\infty}(k,t)} = \frac{1}{C}\Psi(k,t),
\end{equation}
where $E_{\infty}(k,t)= C\epsilon_{\mathrm{num}}^{2/3}(t)k^{-5/3}$.
The second moment of the squared transfer function is given
by\footnote{ Usually, filter length scales are defined by the second
moment of $\hat{G}$ rather than $\hat{G}^{2}$. Computationally,
however, it is preferable to use the square of the transfer function.}
\begin{equation}
  M^{(2)}[\hat{G}_{\mathrm{eff}}^{2}] = 
  \int_{0}^{\infty}k^{2}\hat{G}_{\mathrm{eff}}^{2}(k,t)\dd k.
\end{equation}
Of course, the numerically computed spectrum function $E(k,t)$ does not
conform with the Kolmogorov law at wave numbers comparable to
$k_{0}=2\pi/L$, because of energy injection on the largest length
scales. However, the contribution of these wave numbers to the above
integral is small due to the factor $k^{2}$.  Thus, we will ignore
this error. 

Discretising the transfer function in the fashion outlined
in section~\ref{sc:spectrum} and cutting off at the wavelength
$\pi/\Delta$, the following approximation to the second moment is
obtained:
\begin{equation}
  \label{eq:snd_momt}
  M^{(2)}[\hat{G}_{\mathrm{eff}}^{2}] \simeq 
  \frac{1}{C}\left(\frac{2\pi}{\alpha L}\right)^{3}
    \int_{0}^{N/2}\dd(\alpha\tilde{k})\sum_{n}\Psi_{n}(t)
    \frac{\tilde{\mu}_{n}}{4\pi}\delta[\alpha(\tilde{k}-\tilde{k}_{n})].
\end{equation}
The second moment of the filter transfer function has the dimension of
inverse length cubed. Hence, a length scale is given by
$(M^{(2)}[\hat{G^{2}}])^{-1/3}$ and is customarily normalised with
respect to the second moment of the sharp cutoff filter. The transfer
function of this filter is given by
$\hat{G}_{\Delta}(k)=\theta(\pi/\Delta-k)$, and the second moment is
$M^{(2)}[\hat{G}_{\Delta}^{2}]=(\pi/\Delta)^{3}/3$. Setting
$M^{(2)}[\hat{G}_{\mathrm{eff}}^{2}]=(\pi/\Delta_{\mathrm{eff}})^{3}/3$,
the \emph{effective filter scaling factor} of the PPM is therefore
estimated to be
\begin{equation}
  \label{eq:num_scale}
  \beta = \frac{\Delta_{\mathrm{eff}}}{\Delta} =
  \frac{N}{2}\left[\frac{3}{C}\sum_{n=1}^{n_{\mathrm{c}}}
                   \Psi_{n}(t)\frac{\tilde{\mu}_{n}}{4\pi}\right]^{-1/3},
\end{equation}
where
$n_{\mathrm{c}}=\max\{n|\tilde{k}_{n}\le\tilde{k}_{\mathrm{c}}\}$.

We calculated $\beta$ for each simulation at several instants of
time. The obtained values listed in the tables~\ref{tb:num_nrsl},
\ref{tb:num_nrh75} and~\ref{tb:num_nrh20} demonstrate that
$\Delta_{\mathrm{eff}}\approx 1.6\Delta$ for fully developed
turbulence.  Only for the lowest Mach number, $V/c_{0}\approx 0.084$,
the scaling factor $\beta\approx 1.8$ was found in the stationary
regime. Consequently, the effect of numerical dissipation appears to
become more pronounced for decreasing Mach number. Of course, the
dependence of $\beta$ on the numerical resolution should be
investigated too. Changes can be expected towards lower resolution,
because the energy-containing and the dissipation subrange will
increasingly overlap. At higher resolution, on the other hand, $\beta$
should asymptotically approach a value independent of
$N$. Unfortunately, the validation of this conjecture would require an
undue amount of computational resources.

In fact, $\Delta_{\mathrm{eff}}$ is much smaller than the length scale
of maximum dissipation, $l_{\mathrm{p}}$, which is given by the
maximum of $\tilde{k}_{n}^{2}\tilde{E}_{n}(t)\propto
\tilde{k}_{n}^{1/3}\tilde{\Psi}_{n}(t)$. For fully developed
turbulence, the peak of dissipation was found to be located close to
the second maximum of $\tilde{\Psi}_{n}(t)$, with a typical value
$l_{\mathrm{p}}\approx 0.065 L\approx 9\Delta$ (see tables~\ref{tb:num_nrsl},
\ref{tb:num_nrh75} and~\ref{tb:num_nrh20}). Thus, $l_{\mathrm{p}}$ can
be considered as the characteristic length scale of the bottleneck
effect. The morphology of the flow on length scales $l\gtrsim
l_{\mathrm{p}}$ is illustrated in figure~\ref{fg:vort3d1620}. Shown
are the isosurfaces of vorticity which correspond to the $97\,\%$ level of 
the probability distribution function. In
order to suppress velocity fluctuations on scales smaller than
$l_{\mathrm{p}}$, a Gaussian filter of characteristic length
$10\Delta_{\mathrm{eff}}\approx 2l_{\mathrm{p}}$ was applied to the
flow realisation at time $\tilde{t}=4.0$.  The characteristic Mach
number is $V/c_{0}=0.66$ and the spectral weight $\zeta=3/4$. The
smoothing of the velocity field over a length
$2\Delta_{\mathrm{eff}}\approx l_{\mathrm{p}}/3$ results in the
visualisation shown in figure~\ref{fg:vort3d324}. Although the length
scales smaller than $l_{\mathrm{p}}$ are subject to significant
numerical dissipation, there is nevertheless a great wealth of
substructure on these scales.  It becomes clear that the tube-like
structures in figure~\ref{fg:vort3d1620} are actually concentrations
of smaller vortices. The magnitude of the flow velocity is
colour-coded in the plots. It appears that a significant fraction of
high velocity fluctuations is present on the length scales $l\lesssim
l_{\mathrm{p}}$ in the bottleneck subrange of the turbulence energy
spectrum.

\begin{figure}[thb]
  \begin{center}
    \vspace{100mm}
    \texttt{This figure is not available in the e-print version.}
    \caption{ Isosurfaces of vorticity for the same flow realisation as in figure 6,
      however, with the velocity field smoothed on the length scale $10\Delta_{\mathrm{eff}}=16.2\Delta$,
      which is about the length scale of peak dissipation. }
    \label{fg:vort3d1620}
  \end{center}
\end{figure}

\begin{figure}[htb]
  \begin{center}
    \vspace{100mm}
    \texttt{This figure is not available in the e-print version.}
    \caption{ Isosurfaces of vorticity for developed turbulence in the simulation
      with $V/c_{0}=0.66$ and $\zeta=0.75$. The velocity field was smoothed
      with a Gaussian filter of characteristic length $2\Delta_{\mathrm{eff}}=3.24\Delta$.
      The colour coding corresponds to the dimensionless velocity $v/V$. }
    \label{fg:vort3d324}
  \end{center}
\end{figure}

\section{Conclusion}

The computation of turbulence energy spectrum functions reveals
several important properties of turbulence in numerical simulations
with the PPM: Firstly, the range of length scales approximately
satisfying Kolmogorov scaling, secondly, characteristic scales
associated with numerical dissipation, and thirdly, the so-called
bottleneck effect, i.~e., an excess of kinetic energy in modes
affected by dissipation.

For the simulations of forced isotropic turbulence discussed in this
paper, there is only a marginal inertial subrange. Nevertheless, we
were able to estimate the Kolmogorov constant for samples of the flow
in the nearly statistically stationary regime to be $C\approx 1.7$,
which agrees very well with published results from more elaborate
simulations. Only in the case of a simulation with Mach number of
order unity and forcing dominated by dilatational components, a
significantly smaller value of $C$ was found.  On the basis of the
notion of an implicit filter, we define the length scale
$\Delta_{\mathrm{eff}}$ specifying the characteristic length of
smoothing due to the numerical scheme. The ratio
$\beta=\Delta_{\mathrm{eff}}/\Delta$ appears to be fairly universal
for statistically stationary turbulence.  In the case of the PPM,
$\beta\approx 1.6$.  Furthermore, it was demonstrated that the
computed spectra exhibit a pronounced bottleneck effect. The
corresponding peak is higher than in spectra obtained from simulations
with spectral methods. The magnitude of the bottleneck effect appears
to be similar to what is obtained in simulations with hyperviscosity.

  The parameter $C_{\epsilon}$ specifying the dimensionless rate of
numerical dissipation assumes values of about $0.5$ for developed
turbulence. This is consistent with the time-averaged asymptotic value
in the limit of large Reynolds numbers calculated from simulations of
incompressible turbulence. From the mean rate of numerical
dissipation, one can also compute an equivalent Smagorinsky
length. The ratio of this length to $\Delta_{\mathrm{eff}}$ is very
close to the analytical value of the Smagorinsky constant
$C_{\mathrm{S}}$. This result supports the presumed statistical
equivalence of numerical dissipation in simulations with the PPM and
the subgrid scale energy transfer in proper LES.

In essence, the presented results suggest that the application of the
PPM in fluid dynamical simulations yields a fair numerical
representation of turbulent flows, for which crucial statistical
parameters are in good agreement with the results obtained with more
accurate methods.

\section{Acknowledgements}

The turbulence simulations were run on the Hitachi SR-8000 of the
\emph{Leibniz Computing Centre} in Munich, and the post-processing of
the data was performed on the IBM p690 of the \emph{Computing Centre
of the Max-Planck-Society} in Garching, Germany. We thank M. Reinecke,
who was very helpful with technical advice.  For the computation of
the turbulence energy spectra, the \textsl{FFTW} implementation of the
the fast Fourier transform algorithm was utilised
\cite{FriJohn98}. The research of W. Schmidt was in part supported by
the priority research program \emph{Analysis and Numerics for
Conservation Laws} of the Deutsche Forschungsgesellschaft. Moreover,
W.~Schmidt and J.~C.~Niemeyer were supported by the Alfried Krupp
Prize for Young University Teachers of the \emph{Alfried Krupp von
Bohlen und Halbach Foundation}.

\end{document}